\documentclass[preprint,prb,aps]{revtex4}
\usepackage{amsmath,amssymb}
\usepackage{graphicx}
\usepackage{dcolumn}
\usepackage{bm,cases}

\begin{document}
\title[]{\bf Removal of ordering ambiguity for a class of position dependent mass quantum systems with an application to the quadratic  Li$\acute{e}$nard type nonlinear oscillators}

\author{V.~Chithiika Ruby, V.~K.~Chandrasekar$^{*}$, M~Senthilvelan and M.~Lakshmanan}

\address{Centre for Nonlinear Dynamics, School of Physics,
Bharathidasan University, Tiruchirapalli - 620 024, India.}
\address{${}^{*}$Centre for Nonlinear Science and Engineering, School of Electrical and Electronics
Engineering, SASTRA University, Thanjavur - 613 401, India }

\begin{abstract}
We consider the problem of removal of ordering ambiguity in position dependent mass quantum systems characterized by a generalized position dependent mass Hamiltonian which generalizes a number of  Hermitian as well as non-Hermitian ordered forms of the Hamiltonian. We implement  point canonical transformation method to map one-dimensional time-independent  position dependent mass  Schr\"{o}dinger equation endowed with potentials onto constant mass counterparts which are considered to be exactly solvable. We observe that a class of mass functions and the corresponding potentials  give rise to solutions that do not depend on any particular ordering,  leading to the removal of ambiguity in it. In this case, it is imperative that the ordering is Hermitian. For non-Hermitian ordering we show that the class of systems can also be exactly solvable and are also shown to be iso-spectral using suitable similarity transformations. We also discuss the normalization of the  eigenfunctions obtained  from both Hermitian and non-Hermitian orderings.   We illustrate the technique with  the quadratic  Li$\acute{e}$nard type nonlinear oscillators, which admit position dependent mass Hamiltonians. 

\end{abstract}


\maketitle

\section{Introduction}
Position-dependent mass (PDM) quantum systems, which especially find valuable applications in 
condensed matter physics \cite{apps}, require ordering between momentum and mass operators in 
the kinetic energy term and also require appropriate modification in the boundary conditions 
since some mass functions are not continuous \cite{bou}. The ordering may be Hermitian or non-Hermitian
since non-Hermiticity in certain situations can also allow the possibility of getting real energy eigenvalues \cite{non-herm, bend}. Weyl ordering \cite{weyl_pdms, weyl_ord}, von Roos ordering \cite{von}, Li and Kuhn ordering \cite{lik} and 
Zhu and Kroemer  ordering \cite{zhu} are some specific procedures that correspond to the Hermitian 
construction of the associated quantum Hamiltonian. Hence different orderings
may add different terms to the potential of the system and the resultant potentials are generally known as effective potentials. This causes certain ambiguity in the usage of ordering in a particular situation.
The ordering problem has attracted many researchers to  search for the best possible
choices of kinetic energy operators for quantum systems with position-dependent
effective mass forms. The von Roos ordering has been considered to be the most general one. 
Recently, it has been proved that the von Roos ordering is not the most general one by 
Trabelsi et al \cite{Trabelsi}  who proposed a general formulation of the kinetic energy 
operator with PDM under which von Roos ordering comes as one of the possibilities. They have also presented a complete classification of the operator. 

Many exactly solvable quantum systems with mass depending on the position have emerged  while studying the quantum dynamics of certain classical nonlinear oscillators. Most of such nonlinear oscillators belong to the quadratic Li\'{e}nard type nonlinear oscillators, for example Mathews-Lakshmanan (ML) oscillator \cite{pmm} and its generalizations \cite{midya} and extensions \cite{carinena, barnana}. Other classes of nonlinear oscillators have also been studied in the context of PDM problem \cite{chith_pt, guha1, guha2}. 

The time-independent generalized Schr\"{o}dinger equation corresponding to the PDM quantum systems may also be  solved by implementing the methods applied to the Schr\"{o}dinger equation corresponding to the constant mass systems. Point canonical transformation method, which relates the PDM and constant mass systems, is a widely used method in this direction \cite{quesne}. 

An important goal in the study of the quantum dynamics of  PDM problem is to overcome the ambiguity problem in ordering, that is to identify effective potentials resulting from all possible choices of ordering  which unambiguously possess the same solutions. In the present work, we consider a generalized $2 N$-parameter kinetic energy operator which  unifies all types of Hermitian and non-Hermitian orderings and investigate the effect of ambiguity in the dynamics of systems endowed with one-dimensional potentials. The associated Hamiltonian of  the generalized  kinetic energy operator which is as such non-Hermitian can become Hermitian on either applying a specific condition on the ordering parameters or through a similarity transformation.  In this work, we start our analysis with the generalized Hermitian Hamiltonian. Since point canonical transformation (PCT) maps the constant mass problem onto the PDM problem, we can employ the reciprocal PCT to the Schr\"{o}dinger equation of the generalized PDM Hermitian Hamiltonian  to obtain a constant-mass Schr\"{o}dinger equation giving rise to the same energy spectrum. The constant mass Schr\"{o}dinger equation will then be obtained in a new coordinate variable and can be considered to be associated with  an exactly solvable potential. It is now simple to derive the PDM Schr\"{o}dinger equation wavefunctions from the knowledge of the constant-mass ones. We identify two possible ways in which the mapping from a generalized variable mass Schr\"{o}dinger equation to constant mass Schr\"{o}dinger equation can be made: (i) keep the mass function arbitrary, but fix the ordering parameters, or (ii) allow the ordering parameters to take arbitrary values but fix the mass function. 

We begin with the choice where the mass is an arbitrary function.  Here the generality of the Hermitian Hamiltonian is lost and it gets reduced to a particular form. In the literature this case is studied using von Roos ordering for different exactly solvable potentials such as the linear harmonic oscillator \cite{pdmh}, the generalized isotonic oscillator \cite{pdmi} and other generalized potentials \cite{guha1, guha2} admitting exceptional orthogonal  polynomials.   In the case of the second choice, where the ordering parameters are arbitrary, the mass function takes a particular class of forms. The class of allowed mass functions in this case is derived explicitly. The PDM systems associated with the class of mass functions  admit eigenfunctions which are free from ordering parameters. It is worth mentioning that the ordering ambiguity is removed in the study of PDM systems when they associate with a class of mass functions and subjected  to Hermitian ordering. 

Since the non-Hermitian ordered Hamiltonian can be related with its Hermitian counterpart through a similarity transformation, we can obtain the solutions of the associated Schr\"{o}dinger equation of the non-Hermitian Hamiltonian from that of Hermitian Hamiltonian. In case of the non-Hermitian ordering, we also  consider both the choices, that is (i) mass is arbitrary and (ii) the ordering parameters are arbitrary. In the first choice, that is  the mass is arbitrary, the $2 N$-parameter PDM non-Hermitian Hamiltonian reduces to one parameter Hamiltonian which admits eigenfunctions including the ordering parameter.  For the second choice,  the $2 N$  ordering parameters are arbitrary and the mass function is fixed. Hence, the $2 N$-parameter PDM  systems associated with the class of mass functions  admit eigenfunctions which also include the ordering parameters and so the ambiguity in ordering  is present.  However, in both Hermitian and non-Hermitian orderings the energy spectrum remains the  same. So it is possible to obtain a mixed class of isospectral Hamiltonians, that is a mixed class of Hermitian and non-Hermitian Hamiltonians admitting the same energy spectrum.  The general study is illustrated  with the quantum versions of two specific forms  of quadratic Li\'enard  type nonlinear oscillators discussed in Ref. \cite{tiwari}.  

The plan of the paper is as follows. In the following section, we discuss about the newly proposed general form of PDM kinetic energy operator and its classification and the problem of ordering ambiguity therein. In Sec. III, we study the quantum solvability of Hermitian  Hamiltonian systems associated with  position dependent mass potentials using PCT and isolate specific classes of potentials for which the ordering problem can be removed. In Sec. IV, we extend our study to non-Hermitian Hamiltonian systems. Next, in Sec. V,  we illustrate the proposed method with the quadratic Li\'{e}nard type nonlinear oscillators whose Hamiltonians are position-dependent mass ones. Finally in section VI, we summarize our results. 

\section{Ordering ambiguity}
In the literature, the kinetic energy operator $\hat{T}$, which consists of position-dependent mass $m(x)$ and momentum $\hat{p}$, is expressed in  numerous  ways in different situations \cite{Trabelsi, newke}. 
We point out a few: 
\begin{enumerate}
\item[(i)] Ben Daniel and Duke (BDD) proposed a form \cite{bd}, ${\displaystyle \hat{T} = \frac{1}{2} \hat{p} \frac{1}{m} \hat{p}}$, 
\item[(ii)] Gora and Williams (GW)'s proposal \cite{gw} is ${\displaystyle \hat{T} = \frac{1}{4}\left[\frac{1}{m} \hat{p}^2 +  \hat{p}^2 \frac{1}{m}\right]}$,
\item[(iii)]  Zhu and Kroemer's (ZK) way of ordering \cite{zhu} of ${\displaystyle \hat{T} = \frac{1}{2}\left[\frac{1}{\sqrt{m}} \hat{p}^2\frac{1}{\sqrt{m}}\right]}$, 
\item[(iv)] Weyl ordered form of kinetic energy operator \cite{weyl_ord} is  ${\displaystyle \hat{T} = \frac{1}{6}\left[\frac{1}{m} \hat{p}^2 +  \hat{p} \frac{1}{m} \hat{p} +  \hat{p^2} \frac{1}{m}\right]}$, 
\item[(v)] von Roos proposed a two-parameter general ordering \cite{von} as ${\displaystyle \hat{T} = \frac{1}{4}\left[ m^{\alpha}\hat{p} m^{\beta} \hat{p} m^{\gamma}  + m^{\gamma}\hat{p} m^{\beta} \hat{p} m^{\alpha}\right]}$, where $\alpha + \beta + \gamma = -1$, 
\item[(vi)] Li and Kuhn (LK) ordering and Morrow and Brownstein (MB) ordering  are special cases of von Roos's ordering \cite{lik, morrow},
\item[(vii)] a more general form than von Roos's symmetric ordering  by taking into account of Weyl ordering, ${\displaystyle \hat{T} = \frac{1}{4(a+1)}\left[a\left(\frac{1}{m} \hat{p}^2 +  \hat{p}^2 \frac{1}{m} \right) \right.}$ ${\left. + m^{\alpha}\hat{p} m^{\beta} \hat{p} m^{\gamma} + m^{\gamma}\hat{p} m^{\beta} \hat{p} m^{\alpha}\right]}$, was proposed in \cite{dutra}. Here, the arbitrary parameter $a$ can be tuned to generate different orderings,  including Weyl ordering.  
\end{enumerate}
All these orderings are Hermitian. 

\subsection{\bf General form of kinetic energy operator}
An even more general form of kinetic energy operator than the forms available in the literature  has been recently introduced by allowing many number of possible mixtures of the fundamental 
term, $m^{\alpha} \hat{\bf p} m^{\beta} \hat{\bf p} m^{\gamma}$, where $\hat{\bf p}$ is the three dimensional momentum operator. 
It reads as  \cite{Trabelsi}
\begin{eqnarray}
\hat{\bf T} = \frac{1}{2}\sum^N_{i = 1} w_i m^{\alpha_i} \hat{\bf p} m^{\beta_i} \hat{\bf p} m^{\gamma_i}, 
\label{geo}
\end{eqnarray}
where  $N$ is an arbitrary positive integer, and the ordering parameters should satisfy 
the constraint $\alpha_i +\beta_i +\gamma_i = -1,\;i =1, 2, 3, ... N$ and $w_i$'s  are real weights which are summed to be $1$. The above form globally connects all the Hermitian orderings (see items (i)-(vii) mentioned above) and also provides a complete classification of Hermitian and non-Hermitian orderings \cite{Trabelsi}.  The operator $\hat{T}$ in (\ref{geo}) possesses $2 N$ free 
ordering parameters.

The operator (\ref{geo}) is not Hermitian in general  and can be re-expressed as 
\begin{eqnarray}
\hspace{-1cm}\qquad \qquad \hat{\bf T} &=& \frac{1}{2}\hat{\bf p}\frac{1}{m}\hat{\bf p}+(\bar{\gamma} - \bar{\alpha}) \frac{i \hbar}{2} {\boldsymbol{\nabla}} \left( \frac{1}{m}\right). \hat{\bf p} + \frac{\hbar^2}{2}\left[\bar{\gamma} {\boldsymbol{\nabla}}^2 \left(\frac{1}{m}\right) + \overline{\alpha\gamma} \left({\boldsymbol{\nabla}}\frac{1}{m}\right)^2 m \right],
\label{geoa}
\end{eqnarray}
where $\hbar$ is Planck's constant and  $\bar{X}=(\bar{\alpha}, \bar{\beta}, \bar{\gamma})$ denotes the weighted mean value, ${\displaystyle \bar{X} = \sum_{i = 1}^N w_i X_i}$ and  $-\bar{\beta} = 1 + \bar{\alpha}+\bar{\gamma}$. Note that 
${\displaystyle \overline{\alpha \gamma} = \sum_{i = 1}^N w_i \alpha_i \gamma_i}$.  The corresponding Hamiltonian  for a potential $V$ can be written as  
\begin{eqnarray}
\hspace{-0.5cm}\quad \hat{\bf H}_{non} &=& \hat{\bf T} + V \nonumber \\
\hspace{-0.5cm}\quad &=& \frac{1}{2}\hat{\bf p}\frac{1}{m}\hat{\bf p}+(\bar{\gamma} - \bar{\alpha}) \frac{i \hbar}{2} {\boldsymbol{\nabla}} \left( \frac{1}{m}\right). \hat{\bf p} + \frac{\hbar^2}{2}\left[\bar{\gamma} {\boldsymbol{\nabla}}^2 \left(\frac{1}{m}\right) + \overline{\alpha\gamma} \left({\boldsymbol{\nabla}}\frac{1}{m}\right)^2 m \right] + V. \label{nhe}
\end{eqnarray}
One can also rewrite the above Hamiltonian as 
\begin{eqnarray}
\hspace{-1cm}\qquad \hat{\bf H}_{non} &=& \frac{1}{2}\hat{\bf p}\frac{1}{m}\hat{\bf p}+(\bar{\gamma} - \bar{\alpha}) \frac{i \hbar}{2} {\boldsymbol{\nabla}} \left( \frac{1}{m}\right). \hat{\bf p}  + V_{eff}, \label{nhe_new}
\end{eqnarray}
where 
\begin{eqnarray}
 V_{eff} = \frac{\hbar^2}{2}\left[\bar{\gamma} {\boldsymbol{\nabla}}^2 \left(\frac{1}{m}\right) + \overline{\alpha\gamma} \left({\boldsymbol{\nabla}}\frac{1}{m}\right)^2 m \right] + V, \label{eff_non} 
\end{eqnarray}   
is known as effective potential. In Eqs. (\ref{nhe}) and (\ref{nhe_new}), 
the subscript $non$ in $\hat{H}_{non}$ implies that the Hamiltonian is non-Hermitian. 

\subsection{\bf Hermitian ordering}
The term proportional to ${\displaystyle {\boldsymbol{\nabla}} \left( \frac{1}{m}\right). \hat{\bf p} }$ is responsible for the Hamiltonian (\ref{nhe}) or (\ref{nhe_new}) being non-Hermitian. Hence, the 
removal of this term makes the operator $\hat{\bf H}_{non}$ to be Hermitian. It can be achieved by applying either 
\begin{itemize}
\item[(i)] a condition $\bar{\alpha} = \bar{\gamma}$, or  
\item[(ii)]  when $\bar{\alpha} \neq \bar{\gamma}$, a similarity transformation which relates the non-Hermitian Hamiltonian ($\hat{\bf H}_{non}$) to its Hermitian counterpart ($\hat{\bf H}_{her}$).  
\end{itemize} 

\subsubsection{\bf Method (i)  ${\bf \bar{\alpha} = \bar{\gamma}}$} 
On applying the condition, $\bar{\alpha} = \bar{\gamma}$, the coefficient of the term ${\displaystyle {\boldsymbol{\nabla}} \left( \frac{1}{m}\right). \hat{\bf p} }$ in (\ref{nhe}) vanishes which results in the Hermitian Hamiltonian ($\hat{\bf H}_{her}$) as    
 \begin{eqnarray}
\hat{\bf H}_{her} &=& \frac{1}{2}\hat{\bf p}\frac{1}{m}\hat{\bf p} + \frac{\hbar^2}{2}\left[\bar{\gamma} {\boldsymbol{\nabla}}^2 \left(\frac{1}{m}\right) + \overline{\alpha\gamma} \left({\boldsymbol{\nabla}}\frac{1}{m}\right)^2 m \right] + V  \nonumber \\ 
                 &=& \frac{1}{2}\hat{\bf p}\frac{1}{m}\hat{\bf p} + V_{eff}. \label{he}
\end{eqnarray}
Here the number of ordering parameters appearing in $\hat{\bf H}_{non}$ is reduced to $(2 N - 1)$ parameters due to the condition ${\bar{\alpha} = \bar{\gamma}}$. 

To illustrate the above, in the following we derive the von Roos ordering which is Hermitian.  
Considering  $N = 2, \; w_1 = w_2 = \frac{1}{2}$ in expression (\ref{geo}) and also implementing the Hermiticity condition, $\bar{\alpha} = \bar{\gamma}$, gives $\alpha_1 + \alpha_2 = \gamma_1 + \gamma_2$. We also have the additional conditions $\alpha_1 + \beta_1 + \gamma_1 = -1$, and $\alpha_2 +\beta_2 +  
\gamma_2 = -1$. One of the possible solutions of  these conditions is $\alpha_1 = \gamma_2$, $\alpha_2 = \gamma_1$ and $\beta_1 = \beta_2$, leading to von Roos ordering of  (\ref{geo}) as   
\begin{eqnarray}
\hspace{-1cm} \qquad \hat{\bf T} = \frac{1}{4}\left[ m^{\alpha_1}\hat{\bf p} m^{\beta_1} \hat{\bf p} m^{\gamma_1} + m^{\gamma_1}\hat{\bf p} m^{\beta_1} \hat{\bf p} m^{\alpha_1}\right].  
\label{von}
\end{eqnarray}
The corresponding Hamiltonian from (\ref{he}) can be expressed as 
 \begin{eqnarray}
\hspace{-1cm}\quad \quad \hat{\bf H}_{her} &=& \frac{1}{2}\hat{\bf p}\frac{1}{m}\hat{\bf p} + \frac{\hbar^2}{2}\left[\left(\frac{\alpha_1 + \gamma_1}{2}\right) {\boldsymbol{\nabla}}^2 \left(\frac{1}{m}\right) + {\alpha_1\gamma_1} \left({\boldsymbol{\nabla}}\frac{1}{m}\right)^2 m \right] + V.  \label{vhe}
\end{eqnarray}
It also shows that the von Roos ordering is not the most general form \cite{Trabelsi}.

\subsubsection{\bf Method (ii)  Similarity transformation} 
The non-Hermitian Hamiltonian ${ \hat{\bf H}_{non}}$ given by (\ref{nhe_new}) can be related to the Hermitian Hamiltonian ${\hat{\bf  H}_{her}}$  by performing the transformation 
\begin{equation}
\hat{\bf H}_{her}  = m^{\eta} {\bf \hat{H}}_{non} m^{-\eta},  \label{heta} 
\end{equation}
which yields
\begin{eqnarray}
\hspace{-0.1cm}\;m^{\eta}\hat{\bf H}_{non}  m^{-\eta} &=& \frac{1}{2}\hat{\bf p}\frac{1}{m}\hat{\bf p}+(\bar{\gamma} - \bar{\alpha} -  2 \eta) \frac{i \hbar}{2}  {\boldsymbol{\nabla}} \left( \frac{1}{m}\right). \hat{\bf p} + \frac{\hbar^2}{2} \left[(\bar{\gamma}-\eta) {\boldsymbol{\nabla}}^2 \left(\frac{1}{m}\right) \right. \nonumber \\ 
\hspace{-0.1cm}& & \hspace{3cm}\left.+  (\overline{\alpha\gamma} -  \eta (\eta -\bar{\gamma} + \bar{\alpha}))  \left({\boldsymbol{\nabla}}\frac{1}{m}\right)^2 m \right]  + V.   
\label{geham_he}
\end{eqnarray}
The above form becomes Hermitian if $2 \eta = \bar{\gamma} - \bar{\alpha}$ and so that we have 
\begin{eqnarray}
\hspace{-1cm}\hat{\bf H}_{her} &=& \frac{1}{2}\hat{\bf p}\frac{1}{m}\hat{\bf p} + \frac{\hbar^2}{2}\left[\left(\frac{\bar{\alpha}+\bar{\gamma}}{2}\right) {\boldsymbol{\nabla}}^2 \left(\frac{1}{m}\right) + \left(\overline{\alpha\gamma} + \frac{1}{4} (\bar{\gamma} - \bar{\alpha})^2\right)  \left({\boldsymbol{\nabla}}\frac{1}{m}\right)^2 m\right]+ V, 
\label{geham_he1}
\end{eqnarray}
where now the effective potential takes the form 
\begin{eqnarray}
\hspace{-0.5cm}\quad V_{eff} &=& V +  \frac{\hbar^2}{2}\left[\left(\frac{\bar{\alpha}+\bar{\gamma}}{2}\right) {\boldsymbol{\nabla}}^2 \left(\frac{1}{m}\right)+ \left(\overline{\alpha\gamma} + \frac{1}{4} (\bar{\gamma} - \bar{\alpha})^2\right)   \left({\boldsymbol{\nabla}}\frac{1}{m}\right)^2 m\right]. 
\label{eff2}
\end{eqnarray}
Note that the number of ordering parameters present in $\hat{\bf H}_{non}$ can be preserved in 
the transformed Hermitian Hamiltonian, $\hat{\bf H}_{her}$, obtained in (\ref{geham_he1}) since 
$\bar{\alpha} \neq \bar{\gamma}$. On substituting $\bar{\alpha} = \bar{\gamma}$ in  (\ref{geham_he1}) we can obtain the form (\ref{he}). Hence the Hamiltonian (\ref{geham_he1}) can be considered to be a general Hermitian ordered form of Hamiltonian (\ref{nhe}) or (\ref{nhe_new}) and so, in this work, we consider this Hermitian ordered form of the Hamiltonian  (\ref{geham_he1}) to study the quantum sovability of PDM potentials.   

To illustrate the transformation from non-Hermitian Hamiltonian to Hermitian Hamiltonian, we consider a simple form of kinetic energy operator ${\hat{T}}$ given in (\ref{geo}) by considering  $N = 1$ and so  $ w_1 = 1$, namely 
\begin{eqnarray}
\hspace{-0.5cm}\;\hat{T} = \frac{1}{2} m^{\alpha_1} \hat{\bf p} m^{\beta_1} \hat{\bf p} m^{\gamma_1}, 
\end{eqnarray}              
whose corresponding Hamiltonian is, 
\begin{equation}
\hat{\bf H}_{non} = \frac{1}{2} m^{\alpha_1} \hat{\bf p} m^{\beta_1} \hat{\bf p} m^{\gamma_1} + V, \qquad \quad \alpha_1 \neq \gamma_1. \label{ne1}
\end{equation} 
This non-Hermitian Hamiltonian can be transformed to Hermitian Hamiltonian through the similarity transformation  (\ref{heta}), 
that is 
\begin{equation}
\hat{\bf H}_{her} = m^{\eta} \hat{\bf H}_{non} m^{-\eta} =  \frac{1}{2} m^{(\alpha_1 + \gamma_1)/2} \hat{\bf p} m^{\beta_1} \hat{\bf p} m^{(\alpha_1 + \gamma_1)/2} + V,  \label{ne1}
\end{equation} 
where $\eta = \frac{\gamma_1 - \alpha_1}{2}$. 

The main motivation of the present work is to suggest a straightforward method of removal of ordering  ambiguity in the PDM problem. We restrict our attention to one-dimensional potentials only due to their simplicity and enormous applications. In the literature, the problem has been studied for certain exactly solvable one-dimensional potentials with the help of supersymmetric quantum mechanics within the framework of von Roos ordering \cite{gonul}. We implement the PCT technique and will obtain explicit forms for a class of one-dimensional exactly solvable potentials which are free from ordering ambiguity. To make the study more general than the available works in the literature, we consider the one-dimensional version of the kinetic energy operator ordered as in (\ref{geoa}) since at present it is considered to be the most general form.

We consider both the non-Hermitian and Hermitian ordered forms of position dependent mass Hamiltonian and study their solvability using the PCT method. In the following section, we consider the Hermitian Hamiltonian of the form (\ref{geham_he1}). We take up the case of non-Hermitian ordered Hamiltonian (\ref{nhe}) in the succeeding section $4$. 

\section{Quantum solvability of Hermitian Hamiltonian}
\label{secher}
We consider  the one-dimensional version of Hermitian Hamiltonian  (\ref{geham_he1}) 
for a potential $V(x)$, 
\begin{eqnarray}
\hspace{-0.1cm}\hat{H}_{her} &= &\frac{1}{2}\hat{p}\frac{1}{m}\hat{p} + \frac{\hbar^2}{2}\left[\left(\frac{\bar{\alpha}+\bar{\gamma}}{2}\right) \frac{d^2}{d x^2}\left(\frac{1}{m}\right) + \left(\overline{\alpha\gamma}+ \frac{1}{4} (\bar{\gamma} - \bar{\alpha})^2\right)  \left(\frac{d}{dx}\left(\frac{1}{m}\right)\right)^2 m\right]\nonumber \\ 
& & \hspace{11cm}+ V(x).  
\label{geham_heo}
\end{eqnarray}
Since $[\hat{x}, \hat{p}] = i \hbar$, we assume the coordinate representation 
${\displaystyle \hat{p} = - i \hbar \frac{d}{d x}}$. 

We approach the non-ambiguity problem with the aim of obtaining the solutions for the system (\ref{geham_heo}) which are independent of the ordering parameters $(\alpha_i, \beta_i, \gamma_i)$. We start with the time-independent Schr\"{o}dinger equation for the Hamiltonian (\ref{geham_heo}),  
\begin{eqnarray}
\hspace{-0.5cm}\psi{''} - \frac{m{'}}{m} \psi{'} + 
\left(\left(\frac{\bar{\alpha}+\bar{\gamma}}{2}\right)  \frac{m{''}}{m}-\left(\overline{\alpha\gamma}+\bar{\gamma}+\bar{\alpha}+\frac{1}{4}(\bar{\gamma} - \bar{\alpha})^2\right)  \frac{m{'^2}}{m^2}\right) \psi \nonumber \\ 
\hspace{0.5cm}+ \frac{2 m}{\hbar^2}\left(E -V(x)\right)\psi  = 0, 
\label{seg_he}
\end{eqnarray}
where ${\displaystyle ' = \frac{d}{dx}}$. 

To solve Eq. (\ref{seg_he}), we now use the point canonical transformation (PCT) \cite{pdmh, pct}, that is change the independent variable $x$ in  (\ref{seg_he}) to a new variable $g$, defined by the relation  $x = F(g)$, and introduce the transformation, 
\begin{equation}
\psi(x) = m^{d}(x) \phi[g(x)], \label{pct} 
\end{equation}
where $d$ is an arbitrary parameter. Defining $V(x) = V(F(g)) \equiv U(g)$, Eq. (\ref{seg_he}) can be 
rewritten as 
\begin{eqnarray}
& &\frac{d^2 \phi}{d g^2} + \left[\left(\frac{2 d -1}{g'}\right) \frac{m'}{m} + \frac{g''}{g'^2} \right] \frac{d \phi}{d g}  + \left[\left(d + \frac{\bar{\alpha}+\bar{\gamma}}{2}\right)\frac{1}{g'^2} \frac{m''}{m}  \right. \nonumber \\
& &\hspace{1.5cm}\left.  +  \frac{m'^2}{m^2} \frac{\left(d(d - 2) - \bar{\alpha} -\bar{\gamma} - \overline{\alpha\gamma}-(\bar{\gamma} - \bar{\alpha})^2/4\right)}{g'^2} + \frac{2 m}{g'^2 \hbar^2} (E -U(g)) \right] \phi = 0.  
\label{seg_phi_he}
\end{eqnarray}

Our strategy now is to deduce a constant mass Schr\"{o}dinger equation from (\ref{seg_phi_he}) in the new coordinate variable $g$ corresponding to the potential $U(g)$ with the same energy eigenvalue $E$ of the PDM system (\ref{seg_he}).  It can be achieved by choosing  
\begin{equation}
\qquad g = \int^x \sqrt{m(x')}\;dx',  \label{gm}
\end{equation} 
and expressing all the terms inside the square brackets in (\ref{seg_phi_he}) uniquely in 
terms of $g$ and its  derivatives with respect to $x$ as follows, 
\begin{eqnarray}
\hspace{-1cm} \qquad \frac{d^2 \phi}{d g^2} + (4 d  -1) \frac{g''}{g'^2}\frac{d \phi}{d g} + \left[ (2 d + \bar{\alpha}+\bar{\gamma})\frac{g'''}{g'^3} +  \left(4d\left(d -\frac{3}{2}\right) - 3 \bar{\alpha}- 3 \bar{\gamma}- 4 \overline{\alpha\gamma} \right. \right.\nonumber \\ 
\hspace{-1cm}\qquad  \left.\left.- (\bar{\gamma} - \bar{\alpha})^2\right) \frac{g''^2}{g'^4}+ \frac{2}{\hbar^2} (E -U(g)) \right] \phi = 0.
\label{segp_he}
\end{eqnarray}
Then one can easily remove the term corresponding to the first derivative of $\phi(g)$ in the resultant equation  by choosing $d = \frac{1}{4}$. Consequently (\ref{segp_he}) reduces to 
\begin{eqnarray}
\hspace{-1cm}\qquad\frac{d^2 \phi}{d g^2} + \left[\left(\bar{\alpha} + \bar{\gamma} + \frac{1}{2}\right) \frac{g'''}{g'^3} 
-\left(\bar{\alpha}^2 + \bar{\gamma}^2 -  2 \bar{\alpha} \bar{\gamma} + 3 \bar{\alpha} +3 \bar{\gamma}  + 4 \overline{\alpha\gamma} +\frac{5}{4}\right) \frac{g''^2}{g'^4} + \frac{2}{\hbar^2} (E - U) \right] \phi = 0. \nonumber \\
\label{seg_phi1_he}
\end{eqnarray}

To deduce the constant mass Schr\"{o}dinger equation from (\ref{seg_phi1_he}) without loss of generality, we may demand that 
\begin{eqnarray}
\left(\bar{\alpha} + \bar{\gamma} + \frac{1}{2}\right) \frac{g'''}{g'^3} 
-\left(\bar{\alpha}^2 + \bar{\gamma}^2 -  2 \bar{\alpha} \bar{\gamma} + 3 \bar{\alpha} +3 \bar{\gamma}  + 4 \overline{\alpha\gamma}+\frac{5}{4}\right) \frac{g''^2}{g'^4} = 0.
\label{geq_he}
\end{eqnarray}
Then Eq. (\ref{seg_phi1_he}) reduces to 
\begin{equation}
\qquad \frac{d^2 \phi}{d g^2} + \frac{2}{\hbar^2} (E - U)\phi = 0, 
\label{seg_phi3_he}
\end{equation}
which is indeed  in a form free from any variable mass dependent term. 
From equation (\ref{seg_phi3_he}), we infer that if we provide an exactly solvable potential, $U(g)$, 
with eigenvalues $E_n,\;\;n=0, 1, 2, 3, ...$ and associated normalized eigenfunctions $\phi_{n}(g)$, 
we can solve the equivalent position dependent mass system (\ref{geham_heo}) and obtain the eigenfunctions, namely  
\begin{equation}
\psi_n (x) = m^{\frac{1}{4}} \phi_{n}(g(x)),\;  n =0,1,2,3,... 
\label{sol}
\end{equation}
with same energy eigenvalues $E_n$, subject to the condition that $\psi_n$'s, $n = 0, 1, 2, 3, ...,$   satisfy all the admissibility and boundary conditions. The eigenfunctions (\ref{sol}) can be normalized as follows,   
\begin{eqnarray}
\langle {\psi}_n|{\psi}_n\rangle &=& \int^{\infty}_{-\infty} m^{1/2} \phi^{*}_{n}(g)\phi_n(g) dx \nonumber \\ 
                                 &=&  \int^{\infty}_{-\infty}  \phi^{*}_{n}(g)\phi_{n}(g) dg = 1,  
\end{eqnarray}
where $g' = \sqrt{m(x)}$. 

To solve Eq. (\ref{geq_he}), one can consider two possibilities: 
\begin{enumerate}
\item $g(x)$ and so the mass function ${\displaystyle m(x) = \left(\frac{d g}{d x}\right)^2 = g'^2}$ is 
arbitrary, while the ordering parameters get fixed. 
\item Ordering parameters are arbitrary, while the form of $g(x)$ and so $m(x)$ get fixed. 
\end{enumerate} 
We consider both the cases separately. 

\subsection{\bf Case (i) Arbitrary mass functions}
\label{case1_he}
In this case, the coefficients in (\ref{geq_he}) must vanish separately for arbitrary mass functions,   
\begin{eqnarray}
 \hspace{-0.5cm}\bar{\alpha} + \bar{\gamma} + \frac{1}{2}&=& 0, \nonumber \\
\mbox{and}\hspace{6cm} & & \nonumber \\
 \hspace{-0.5cm}\bar{\alpha}^2 + \bar{\gamma}^2 -  2 \bar{\alpha} \bar{\gamma} + 3 \bar{\alpha} +3 \bar{\gamma} + 4 \overline{\alpha\gamma}+\frac{5}{4}&=&0. \label{cc}
\end{eqnarray}
Using the condition $\alpha_i+\beta_i+\gamma_i = -1$ in the first equation of (\ref{cc}), we obtain 
\begin{equation}
\bar{\beta} = - \frac{1}{2}. \label{beta} 
\end{equation}
On solving the remaining equation in (\ref{cc}), we obtain  
\begin{equation}
\overline{\gamma^2} = (\bar{\gamma})^2 \qquad \mbox{and}\qquad \overline{\alpha^2} = (\bar{\alpha})^2, \label{ag_he}
\end{equation}
which fix 
\begin{equation}
\overline{\beta^2} = (\bar{\beta})^2.
\label{be_he}
\end{equation} 
This means that the variances in the parameters are zero. It results that all $\gamma_i$'s  are the same and so also $\alpha_i$'s and $\beta_i$'s,  where $i =1, 2, 3, ...N$. 
Hence we can consider  
\begin{equation}
\alpha_i = \alpha,\;\;\beta_i = \beta\;\;\mbox{and}\;\;\gamma_i = \gamma,\;\;i = 1, 2, 3, ...N.  
 \label{abg_he}
\end{equation}

On implementing the result (\ref{abg_he}) in Eqs. (\ref{cc}) and (\ref{beta}), we can obtain the conditions 
\begin{equation}
\alpha = - \frac{1}{2} - \gamma\;\;\mbox{and}\;\;\beta = -\frac{1}{2}. \label{cccc}
\end{equation}
With this choice the Hermitian Hamiltonian (\ref{geham_heo}) takes the form
\begin{eqnarray}
\hat{H}_{her} =  \frac{1}{2} \hat{p} \frac{1}{m} \hat{p} + \frac{\hbar^2}{8 m} \left[\frac{m''}{m} - \frac{7}{4}\left(\frac{m'}{m}\right)^2\right],  
\label{geo_1}
\end{eqnarray}
which can also be re-written as 
\begin{eqnarray}
\hspace{-1cm}\qquad \hat{H}_{her} =  \frac{1}{2} m^{-\frac{1}{4}} \hat{p} m^{-\frac{1}{2}} \hat{p} m^{-\frac{1}{4}}+ V(x). \label{her_he}
\end{eqnarray}

From (\ref{sol}), we get the eigenfunctions of the system (\ref{her_he}), 
\begin{eqnarray}
{\psi}_n (x) &=& m^{1/4} \phi_{n}(g(x)), \hspace{1cm} n =0,1,2,3,..., \nonumber 
\label{sol_her1}
\end{eqnarray}
which do not include ordering parameters. 
Hence we can conclude that the PDM Hamiltonians take only one 
particular form (\ref{her_he}) for the arbitrary mass functions when the ordering is restricted to be  Hermitian \cite{guha1}. 

\subsection{\bf Case (ii) Arbitrary ordering parameters}
\label{case2_he}
Here we consider that the ordering parameters can be arbitrary, so that either 
$\bar{\alpha} + \bar{\gamma} + \frac{1}{2} \neq 0$ or 
$\bar{\alpha}^2 + \bar{\gamma}^2 -  2 \bar{\alpha} \bar{\gamma} + 3 \bar{\alpha} +3 \bar{\gamma} + 4 \overline{\alpha\gamma}+\frac{5}{4} \neq 0$ or both are non-zero.  

In this case, we try to solve (\ref{geq_he}) generally by considering the general case, 
$\bar{\alpha} + \bar{\gamma} + \frac{1}{2} \neq 0$ and 
$\bar{\alpha}^2 + \bar{\gamma}^2 -  2 \bar{\alpha} \bar{\gamma} + 3 \bar{\alpha} +3 \bar{\gamma} + 4 \overline{\alpha\gamma}+\frac{5}{4} \neq 0$. To do so we consider the transformation 
\begin{equation}
x = F(g) \equiv \int u(g) dg \quad \mbox{or} \quad \frac{d g}{d x} = \frac{1}{u(g)}, 
\end{equation} 
where  the function $u(g)$ is to be determined. Then the above equation (\ref{geq_he}) becomes 
\begin{eqnarray}
\hspace{-1cm}\qquad\left(\bar{\alpha} + \bar{\gamma} + \frac{1}{2}\right) \frac{\ddot{u}}{u} 
+\left(\bar{\alpha}^2 + \bar{\gamma}^2 - 2 \bar{\alpha} \bar{\gamma} + 4 \overline{\alpha\gamma} 
-\frac{1}{4}\right)  \frac{\dot{u}^2}{u^2}  = 0,  \nonumber \\ 
\label{ugeq}
\end{eqnarray}
where ${\displaystyle ({}^{.}) = \frac{d}{d g}}$.

The  second-order nonlinear differential equation (\ref{ugeq}) can be transformed into a  first-order equation through a further transformation, ${\displaystyle -\frac{\dot{u}}{u}} = \theta(g)$: 
\begin{eqnarray}
\left(\bar{\alpha} + \bar{\gamma} + \frac{1}{2}\right) \dot{\theta} 
-\left(\bar{\alpha}^2 + \bar{\gamma}^2 - 2 \bar{\alpha} \bar{\gamma}+ \bar{\alpha} + \bar{\gamma} + 4 \overline{\alpha\gamma} +\frac{1}{4}\right)  \theta^2  = 0, 
\label{tgeq}
\end{eqnarray}
which is obviously  a Bernoulli equation that can be transformed to a linear first-order differential  equation with ${\displaystyle \theta = \frac{1}{w}}$ as  
\begin{eqnarray}
\hspace{-1cm} \dot{w} =  c_1, \;\displaystyle \frac{\left(\bar{\alpha}^2 + \bar{\gamma}^2 - 2 \bar{\alpha} \bar{\gamma} + \bar{\alpha}+ \bar{\gamma} + 4\overline{\alpha\gamma} +\frac{1}{4}\right)}{\left(\bar{\alpha} + \bar{\gamma} + \frac{1}{2}\right)}  = c_1. 
\label{wgeq}
\end{eqnarray}
Note that $c_1$ is an arbitrary parameter.   

By solving the resultant equation (\ref{wgeq}), one can obtain 
\begin{eqnarray}
w(g) = c_1 g + c_2, \label{w}
\end{eqnarray}
where $c_2$ is an integration constant. Consequently we use the result (\ref{w}) in the above transformations and obtain 
\small
\begin{eqnarray}
\theta &=& \frac{1}{w} = \frac{1}{c_1 g + c_2} =\frac{\displaystyle s}{g + \frac{c_1}{c_2}} \nonumber \\
 \Rightarrow u(g) &= & c_3 \exp{\left(-\int \theta(g) dg \right)} = \frac{c_3}{\left(g + C_1\right)^{\displaystyle s}},   
\end{eqnarray}
\normalsize
where ${\displaystyle s = \frac{1}{c_1} , \;\; C_1 = \frac{c_2}{c_1}}$ and $c_3$ is a constant of integration. Consequently we have 
\begin{equation}
g' = \frac{d g}{d x} = \frac{1}{u(g)} = C_2 \left(g + C_1\right)^{\displaystyle s}, \label{gp}
\end{equation}
where ${\displaystyle C_2 = \frac{1}{c_3}}$. 

While solving (\ref{seg_phi3_he}), the function $g'(x) = C_2  \left(g + C_1\right)^s$ is fixed and so  
\begin{eqnarray}
 g(x) = \left\{\begin{array}{c}
 C_3\; \exp{\left(C_2 x\right)} - C_1,  \quad\;\mbox{if}\;\;s = 1 \\
  \left[(1 - s) \left(C_2 x + C_3\right)\right]^{\displaystyle \frac{1}{1 - s}} - C_1, \;\mbox{if}\;\;s \neq 1,\\
  \end{array}\right.\label{g}
\end{eqnarray}
where $C_3$ is  a constant of integration. 
Since from (\ref{gm}) $m(x) =  g'^2(x)$, the expressions (\ref{g}) for $g(x)$ explicitly fix the mass function as 
\begin{eqnarray}
 m(x) = \left\{\begin{array}{c}
 {C^2_3\;C^2_2}\;\exp{\left(2 C_2 x\right)},  \quad \mbox{if}\;\;s = 1 \\
  C^2_2 \left[(1 - s) \left(C_2 x + C_3\right)\right]^{\displaystyle \frac{2 s}{1 - s}}, \;\mbox{if}\;\;\;s \neq 1.\\
  \end{array}\right. \label{mass}
\end{eqnarray}
Now we can redefine the constants and express the allowed $g(x)$ as  
\begin{eqnarray}
g(x) = \left\{\begin{array}{c}
 \hspace{-3.4cm} \mu_1\;e^{\mu_2 x} - \mu_3,  \\
   \left(\nu_1 x + \nu_2\right)^{c} - \mu_3, \;\;\;-\infty < c < \infty,\\
  \end{array}\right.\label{gess}
\end{eqnarray}
where $\mu_1 = C_3$, $\mu_2 = C_2,$ $\mu_3 = C_1$, $\nu_1 = (1 - s) C_2$, $\nu_2 = (1 - s) C_1$, ${\displaystyle c = \frac{1}{1-s}}$. 
The corresponding mass functions are given by 
\begin{eqnarray}
 m(x) = \left\{\begin{array}{c}
\;\hspace{-3cm} a_1\;e^{a_2 x},  \\
 (b_1 x + b_2)^{C}, \;-\infty < C < \infty,\\
  \end{array}\right. \label{massess}
\end{eqnarray}
where $a_1 = C^2_3 C^2_2$, $a_2 = 2 C_2,$ and $b_1 = (1-s)\;C^{\frac{1}{s}}_2$, $b_2 =  (1-s)\;C^{\frac{1-s}{s}}_2 C_3$, and  ${\displaystyle C = \frac{2 s}{1-s}}$. 
Every specific value of $C$ yields a particular mass profile (\ref{massess}). Since  $C$ can take infinite set of values, Eq. (\ref{massess}) corresponds to an  infinite set of position dependent mass forms. It is to be  noted that the corresponding kinetic energy operator (\ref{geham_heo}) has $2 N$ parameters which are arbitrary. The method shows that any constant mass exactly solvable potential can be transformed to a class of exactly solvable position dependent mass potentials corresponding to each mass function (\ref{massess}) without worrying about a particular ordering of Hamiltonian of the quantum system, subject to appropriate boundary conditions being satisfied. 

Hence, one can solve the position dependent mass system (\ref{geham_heo}) 
for the choice of $m(x)$ given by (\ref{massess}) and obtain the eigenfunctions (\ref{sol}), namely  
\begin{eqnarray}
\psi_n (x) &=& m^{\frac{1}{4}} \phi_{n}(g(x)), \hspace{0.3cm} n =0,1,2,3,... 
\end{eqnarray}
with the same energy eigenvalues $E_n$ for arbitrary ordering parameters.  

Here we conclude that the above class of mass functions (\ref{massess}) removes the ordering ambiguity in the associated position dependent mass systems (\ref{geham_heo}). 

\subsection{\bf Continuity condition of the eigenfunction (\ref{sol})} 

In the above two cases (i) and (ii), $m(x)$ can be singular for example when $C < 0$ in Eq. (\ref{massess}). Hence one has to determine suitable continuity conditions for the wavefunction. The Hermitian Hamiltonian (\ref{geham_heo}) is consistent with the continuity equation of standard form of position dependent mass system with 
probability current density, 
\begin{equation}
j = \frac{\hbar}{2 i} \left[\psi^* \frac{1}{m} \frac{\partial}{\partial x} \psi - \psi \frac{1}{m} \frac{\partial}{\partial x}\psi^*\right]. \label{cont}
\end{equation} 

Consider the case where mass function $m(x)$ has a discontinuity at $x = 0$, as an example. Then we have to find out the matching conditions, that is how  $\psi$ and ${\displaystyle \frac{\partial \psi}{\partial x}}$ at $x = 0^-$ are related to their values at $x = 0^{+}$, where the indices $-$ and $+$ denote, respectively, the left and right-hand sides of the mass discontinuity  point in $x$ 
\cite{morrow, intertwin}. Starting with the constant mass Schr\"{o}dinger equation (\ref{seg_phi3_he}) 
\begin{eqnarray}
\qquad \qquad \frac{d^2 \phi}{d g^2} + \frac{2}{\hbar^2} (E - U)\phi = 0, \nonumber \hspace{6.35cm} 
\end{eqnarray}
and substituting the solution (\ref{sol})  
\begin{eqnarray}
\qquad \qquad \phi_n(g(x)) = m^{-1/4} \psi_{n}(x), \label{con_phi}
\end{eqnarray}
and also the transformation $g' = \sqrt{m(x)}$, we obtain 
\begin{equation}
\frac{d}{d x}\left(\frac{1}{m} \frac{d \psi}{d x}\right) - \left(\frac{m''}{4 m'^2} - \frac{7}{16}\frac{m'^2}{m^4}\right) \psi + \frac{2}{\hbar^2} (E - V) \psi = 0. \label{he_cc}
\end{equation}
For the continuous potential $U(g(x))$ of  Eq. (\ref{seg_phi3_he}), $\phi(g)$ and ${\displaystyle \frac{d \phi}{d g}}$ should be 
continuous.  Then Eq. (\ref{he_cc}) must also be continuous for the same potential $V(x)\equiv U(g(x))$. One can observe that Eq. (\ref{he_cc}) is nothing but  the time-independent Schr\"{o}dinger for the Hermitian Hamiltonian (\ref{her_he}),  
\begin{eqnarray}
\hat{H}_{her} = \frac{1}{2} m^{-\frac{1}{4}} \hat{p} m^{-\frac{1}{2}} \hat{p} m^{-\frac{1}{4}}+ V(x).  \nonumber 
\end{eqnarray}
For this Hamiltonian, from the continuity of $\phi_{n}(g(x))'s$, using (\ref{con_phi}) we can write the  continuity conditions for $\psi_n'$s as  
\begin{eqnarray}
\mbox{(i)}\hspace{0.8cm} (m^{-1/4}{\psi})_{-} &=& (m^{-1/4}{\psi})_{+}, \nonumber \\
\mbox{(ii)}\;\; \left(m^{-3/4}\frac{d {\psi}}{d x} \right)_{-} &=& \left(m^{-3/4}\frac{d {\psi}}{d x} \right)_{+}.
\end{eqnarray}
 Note that a similar condition was used in the case of von Roos ordering for $\alpha = \beta$ in Ref. \cite{morrow}. 
 
\section{Solvability of non-Hermitian Hamiltonian}
As we have pointed out in Sec. II the non-Hermitian Hamiltonian for a one-dimensional potential $V(x)$, as deduced from Eq. (\ref{nhe}), is 
\begin{eqnarray}
\hspace{-1cm}  \qquad \hat{H}_{non} = \frac{1}{2}\hat{p}\frac{1}{m}\hat{p}+(\bar{\gamma} - \bar{\alpha}) \frac{i \hbar}{2} \frac{d}{dx}\left(\frac{1}{m}\right) 
\hat{p} + \frac{\hbar^2}{2}\left[\bar{\gamma} \frac{d^2}{dx^2}\left(\frac{1}{m}\right) 
 + \overline{\alpha\gamma} \left(\frac{d}{dx}\left(\frac{1}{m}\right)\right)^2 m \right]+ V(x). \nonumber \\
\label{geham_non}
\end{eqnarray}
It can be related to the Hermitian Hamiltonian (\ref{geham_heo}) through the transformation (\ref{heta}), 
\begin{eqnarray}
\hat{H}_{non} = m^{-\eta} \hat{H}_{her} m^{\eta}, \quad  \eta = \frac{\bar{\gamma} - \bar{\alpha}}{2}.  \label{n1}
\end{eqnarray} 
Hence  the solution of the associated one-dimensional time-independent Schr\"{o}dinger equation of the Hamiltoninan (\ref{n1}), namely 
\begin{eqnarray}
 \frac{-\hbar^2}{2 m}\left[\tilde{\psi}{''} + (\bar{\gamma} -  \bar{\alpha} - 1) \frac{m{'}}{m} \tilde{\psi}{'} +  
\left( \bar{\gamma} \frac{m{''}}{m} -(\overline{\alpha\gamma}+2\bar{\gamma})  \times\frac{m{'^2}}{m^2}\right)\tilde{ \psi}\right] + V(x)\tilde{\psi} =  E \tilde{\psi},\; ' = \frac{d}{dx}, \nonumber \\ 
\label{seg_non}
\end{eqnarray} 
can be obtained using the relation, 
\begin{equation}
\tilde{\psi}_n = m^{-\eta} {\psi}_n(x) = m^{\frac{\bar{\alpha} - \bar{\gamma}}{2}} {\psi}_n(x),\;\mbox{where}\;\eta = \frac{\bar{\gamma}-\bar{\alpha}}{2},  \label{sol_non}
\end{equation}
for the same energy eigenvalue $E_n$ of the Hermitian Hamiltonian (\ref{geham_heo}), that is 
$\hat{H}_{her} \psi_n = E_n \psi_n$. On substituting the solution of Hermitian Hamiltonian (\ref{sol}) in (\ref{sol_non}), we can explicitly express the eigenfunctions of the non-Hermitian Hamiltonian as 
\begin{eqnarray}
\tilde{\psi}_n (x) &=& m^{\frac{1}{2}\left(\bar{\alpha} -  \bar{\gamma} + \frac{1}{2}\right)} \phi_{n}(g(x)),  \;\;
n =0,1,2,3,..., 
\label{sol_n2}
\end{eqnarray}
which includes the ordering parameters. Here, $\phi_{n}(g(x))$ are the normalized eigenfunctions of the constant mass Schr\"{o}dinger equation (\ref{seg_phi3_he}). The eigenfunctions (\ref{sol_n2}) may be singular when the mass function $m(x)$ is singular for some values of ordering parameters. It can be avoided by considering, $\bar{\alpha} -  \bar{\gamma} - \frac{3}{2}<0$.  Different choices of ordering parameters result in different effective potentials, which can be identified from (\ref{eff_non}), whose eigenfunctions (\ref{sol_n2}) are also different but the energy eigenvalues are the same. Such potentials are also known as iso-spectral potentials.

Now we discuss the results with the two cases, that is (i) $m(x)$ is arbitrary and (ii) the ordering parameters are arbitrary. 

{\bf Case (i)}
In this case, we keep the mass function as arbitrary which fixes the coefficients in (\ref{geq_he}) as  
\begin{eqnarray}
\alpha = - \frac{1}{2} - \gamma\;\;\mbox{and}\;\;\beta = -\frac{1}{2},  \nonumber 
\end{eqnarray}
which can be obtained from   Eqs. (\ref{cc}), (\ref{beta}) and (\ref{abg_he}). 

With this choice the kinetic energy operator (\ref{geo}) takes the form
\begin{eqnarray}
\hat{T} =  \frac{1}{2} m^{-\frac{1}{2}-\gamma} \hat{p} m^{-\frac{1}{2}} \hat{p} m^{\gamma}. 
\label{geo_1}
\end{eqnarray}
It fixes the one dimensional non-Hermitian Hamiltonian ($\hat{H}_{non}$) to be of the form 
\begin{equation}
\hat{H}_{non} = \frac{1}{2} m^{-\frac{1}{2}-\gamma} \hat{p} m^{-\frac{1}{2}} \hat{p} m^{\gamma}+ V(x).  
\label{herH1}
\end{equation}
Here we infer that the PDM Hamiltonian reduces to one-parameter Hamiltonian (\ref{herH1}) 
for arbitrary form of mass functions when the ordering is considered to be non-Hermitian.

The eigenfunctions of the system (\ref{herH1}) can be  obtained from (\ref{sol_n2}) as 
\begin{equation}
\tilde{\psi}_n (x) = m^{-\gamma} \phi_{n}(g(x)), \;\; n =0,1,2,3,.... 
\end{equation}

{\bf Case (ii)}
Keeping $2 N$ ordering parameters as arbitrary in (\ref{geq_he}) yields the relation (\ref{gp}), 
${\displaystyle g' =  C_2 \left(g + C_1\right)^s}$,  which fixes  $g(x)$ as in (\ref{g}) or (\ref{gess}) and $m(x)$ as given by  (\ref{massess}). For this case the solutions are obtained in (\ref{sol_n2}). Hence, for the mass function (\ref{massess}),  the general non-Hermitian Hamiltonian (\ref{geham_non})  can be solved for any choice of ordering parameters.

\subsection{\bf Normalization and continuity conditions of (\ref{sol_n2})}
In the case of non-Hermitian ordering, $\bar{\alpha}  \neq \bar{\gamma}$, we also obtain real energy eigenvalues. The standard inner product definition, that is $\langle \tilde{\psi}_m|H_{non}\tilde{\psi}_n\rangle = \langle \tilde{\psi}_m|H^{\dagger}_{non} \tilde{\psi}_n\rangle$, has to be modified to prove   the reality of the  energy eigenvalues of the non-Hermitian Hamiltonian \cite{non-her}.  Here  $\langle \tilde{\psi}_m|$ is not dual to $| \tilde{\psi}_m\rangle$ since the associated Hamiltonian $\hat{H}_{non}$ is not Hermitian. 
We first find out the correspondence  between $\hat{H}_{non}$ and its conjugation $\hat{H}^{\dagger}_{non}$.  
Now we consider the relation (\ref{heta}), 
\begin{equation}
\hat{H}_{her} = m^{\eta}\hat{H}_{non} m^{-\eta}, \label{dir3}    
\end{equation}  
and its associated Hermitian conjugation, 
\begin{equation}
\hat{H}^{\dagger}_{her} = m^{-\eta}\hat{H}^{\dagger}_{non} m^{\eta}.  \label{dir4}   
\end{equation}
Since $\hat{H}_{her} = \hat{H}^{\dagger}_{her}$, equating the above equations (\ref{dir3}) and (\ref{dir4}) yields a relation, 
\begin{equation}
 \hat{H}^{\dagger}_{non} = m^{2\eta}\hat{H}_{non} m^{-2\eta}, \;\; \eta = \frac{\bar{\gamma} - \bar{\alpha}}{2}.  \label{dir5}
\end{equation} 
It can be expanded as 
\small
\begin{eqnarray}
\hspace{-0.3cm}\hat{H}^{\dagger}_{non} = \frac{1}{2}\hat{p}\frac{1}{m}\hat{ p}-(\bar{\gamma} - \bar{\alpha}) \frac{i \hbar}{2}  { \frac{d}{dx}} \left( \frac{1}{m}\right). \hat{p}  + \frac{\hbar^2}{2}\left[\bar{\alpha}\frac{d^2}{d x^2} \left(\frac{1}{m}\right) + \overline{\alpha\gamma} \left(\frac{d}{dx}\left(\frac{1}{m}\right)\right)^2 m \right] + V.
\end{eqnarray}
\normalsize
It is instructive to compare   the corresponding form of $\hat{H}_{non}$ given in Eq. (\ref{geham_non}). 
The eigenstates of $\hat{H}^{\dagger}_{non}$ can be found out from the expressions (\ref{sol_n2}) and (\ref{dir5}) as 
\begin{equation}
\tilde{\phi}_n = m^{2 \eta} \tilde{\psi}_n. \label{dir6} 
\end{equation}
We can now show that indeed $|\tilde{\phi}_m\rangle$ is dual to $\langle \tilde{\psi}_m|$. 

To prove that  $\langle \tilde{\phi}_m|$ is  dual to $| \tilde{\psi}_m\rangle$ rather than $\langle \tilde{\psi}_m|$, we first represent the one dimensional time-independent Schr\"{o}dinger equation (\ref{geham_non}) in terms of Dirac's notation as 
\begin{equation}
\hat{H}_{non} |\tilde{\psi}_{n}\rangle = E_n |\tilde{\psi}_{n} \rangle, \nonumber 
\end{equation}
and the inner product as 
\begin{equation}
\langle \tilde{\phi}_n|\hat{H}_{non} |\tilde{\psi}_{n}\rangle = E_n \langle \tilde{\phi}_n|\tilde{\psi}_{n} \rangle.  \label{dir1}
\end{equation}
The adjoint of (\ref{dir1}) is 
\begin{equation}
\langle \tilde{\psi}_n|\hat{H}^{\dagger}_{non}|\tilde{\phi}_{n}\rangle =  E_n \langle \tilde{\psi}_n|\tilde{\phi}_{n}\rangle, \label{dir2}
\end{equation}

On substituting  (\ref{dir5}) in (\ref{dir2}), we can obtain 
\begin{equation}
\langle \tilde{\psi}_n|m^{2\eta} \hat{H}_{non} m^{-2\eta}|\tilde{\phi}_{n}\rangle =  E_n \langle \tilde{\psi}_n|\tilde{\phi}_{n}\rangle, \label{dir7}
\end{equation}
Using (\ref{dir6}) in (\ref{dir2}), we can transform it to 
\begin{equation}
\langle \tilde{\phi}_n| \hat{H}_{non} |\tilde{\psi}_{n}\rangle =  E_n \langle \tilde{\psi}_n|\tilde{\phi}_{n}\rangle.  \label{dir8}
\end{equation}

On equating (\ref{dir2}) with (\ref{dir8}), we can obtain the relation, 
\begin{equation}
\langle \tilde{\phi}_n| \hat{H}_{non} |\tilde{\psi}_{n}\rangle = \langle \tilde{\psi}_n|\hat{H}^{\dagger}_{non}|\tilde{\phi}_{n}\rangle, \label{dir9}
\end{equation}
which in turn proves the reality of the energy spectrum of the non-Hermitian Hamiltonian. 
Similarly from Eqs. (\ref{dir1}) and (\ref{dir8}), 
\begin{equation}
\langle \tilde{\psi}_n | \tilde{\phi}_n \rangle = \langle \tilde{\phi}_n |\tilde{\psi}_n \rangle. 
\end{equation}
Hence it is confirmed that  $\langle \tilde{\phi}_n|$ is dual to $|\tilde{\psi}_n\rangle$.  

Now we evaluate $\langle \tilde{\phi}_n |\tilde{\psi}_n \rangle $, 
\begin{eqnarray}
 \langle \tilde{\phi}_n |\tilde{\psi}_n \rangle &=& \int^{\infty}_{-\infty} \tilde{\phi}^*_n(x) \tilde{\psi}_n dx \nonumber\\
 &=&  \int^{\infty}_{-\infty} m^{2\eta}\tilde{\psi}^*_n(x) \tilde{\psi}_n dx , \qquad \mbox{from}\;\;(\ref{dir6}) \nonumber \\
   &=& \int^{\infty}_{-\infty} m^{2\eta}m^{-2\eta+\frac{1}{2}}\phi^*_n(g(x))\phi_n(g(x)) dx , \;\;\mbox{from}\;\;(\ref{sol_n2}) \nonumber \\
   &=&  \int^{\infty}_{-\infty} \phi^*_n(g(x))\phi_n(g(x)) dg , \;\; \mbox{where}\;\;g' = \sqrt{m(x)}, 
\nonumber \\
 &=&  1. \label{dir10}
\end{eqnarray}
Hence $\psi_n$ is normalized with respect to $m^{2\eta}$ (let it be $\rho$). So 
the inner product can be represented as follows: 
\begin{equation}
\langle \tilde{\phi}_n |\tilde{\psi}_n \rangle = \langle \tilde{\psi}_n |m^{2\eta}\tilde{\psi}_n \rangle = \langle \tilde{\psi}_n |\tilde{\psi}_n \rangle_{\rho}. 
\end{equation} 

To derive the continuity condition for the wavefunction  $\tilde{\psi}$ of $\hat{H}_{non}$, we first  
evaluate $\tilde{\phi}^* \hat{H}_{non} \tilde{\psi}$ from (\ref{geham_non}) as 
\begin{eqnarray}
\hspace{-1cm} \qquad -\frac{\hbar^2}{2}\tilde{\phi}^* \frac{d}{dx}\frac{1}{m}\frac{d \tilde{\psi}}{d x}+(\bar{\gamma} - \bar{\alpha}) 
\frac{\hbar^2}{2} \frac{d}{dx}\left(\frac{1}{m}\right) \tilde{\phi}^* \frac{d \psi}{d x} + \frac{\hbar^2}{2}\left[\bar{\gamma} \frac{d^2}{dx^2}\left(\frac{1}{m}\right) + \overline{\alpha\gamma} \left(\frac{d}{dx}\left(\frac{1}{m}\right)\right)^2 m \right] \nonumber \\ 
\times\tilde{\phi}^* \tilde{\psi} + V(x) \tilde{\phi}^* \tilde{\psi}= E \tilde{\phi}^* \tilde{\psi},
 \label{dir11}
\end{eqnarray}
and also $\tilde{\psi} \hat{H}^{\dagger}_{non} \tilde{\phi}^*$,
\begin{eqnarray}
\hspace{-1cm} \qquad-\frac{\hbar^2}{2}\tilde{\psi}\frac{d}{dx}\frac{1}{m}\frac{d \tilde{\phi}^*}{d x} - (\bar{\gamma} - \bar{\alpha}) \frac{\hbar^2}{2} \frac{d}{dx}\left(\frac{1}{m}\right) \tilde{\psi} \frac{d \tilde{\phi}^*}{d x} + \frac{\hbar^2}{2}\left[\bar{\alpha} \frac{d^2}{dx^2}\left(\frac{1}{m}\right)  + \overline{\alpha\gamma} \left(\frac{d}{dx}\left(\frac{1}{m}\right)\right)^2 m \right] \nonumber \\
\times\tilde{\psi} \tilde{\phi}^* + V(x) \tilde{\psi} \tilde{\phi}^* = E \tilde{\psi} \tilde{\phi}^*. 
\label{dir12}
\end{eqnarray}

On subtracting (\ref{dir12}) from (\ref{dir11}), we can write 
\begin{equation}
\frac{d}{d x}(j) = 0, 
 \label{cc_non} 
\end{equation}
where 
\begin{equation}
j = \frac{\hbar}{2 i}\left[\frac{1}{m}\left(\phi^*\frac{d\psi}{dx}- \psi\frac{d\phi^*}{dx}\right) - (\bar{\gamma} - \alpha) \frac{d}{dx}\left(\frac{1}{m}\right) \phi^* \psi\right]. 
\label{cc_non2}
\end{equation}

Hence Eq. (\ref{cc_non}) is nothing but the continuity equation for the stationary state $\tilde{\psi}$ of 
$\hat{H}_{non}$ and $j$ can be interpretated as the current density. To do so, we substitute 
$\tilde \phi = m^{2 \eta} \tilde{\psi} $ (vide (\ref{dir6})), where $\eta = \frac{\bar{\gamma}-\bar{\alpha}}{2}$, in Eq. (\ref{cc_non2}) and obtain 
\begin{eqnarray}
j = \frac{\hbar}{2 i}m^{2\eta}\left[\frac{1}{m}\left(\psi^*\frac{d\psi}{dx}- \psi\frac{d\psi^*}{dx}\right) \right]. 
\label{cc_non3}
\end{eqnarray}

Now we integrate the probability flux $j$ over all space 
\begin{eqnarray}
\hspace{-1cm}\qquad \int^{\infty}_{-\infty} j dx &=&  \int^{\infty}_{-\infty} m^{2\eta} \left[\tilde{\psi^*} \frac{1}{m} \hat{p}\tilde{\psi} +\tilde{\psi} \left(\frac{1}{m} \hat{p}\tilde{\psi^*}\right)\right] dx \nonumber \\
  & = & \langle \tilde{\psi}|\frac{1}{m} \hat{p} \psi \rangle_{\rho} + \langle \frac{1}{m}\hat{p} \tilde{\psi}|\psi\rangle_{\rho} \equiv \langle\frac{1}{m}\hat{p}\rangle. 
  \label{cc_non4}  
\end{eqnarray}
This is consistent with the definition of current density of constant mass eigenfunction $\phi(g(x))$ \cite{schiff}.

\section{General Li\'{e}nard type nonlinear oscillators and their quantization}
In this section, we illustrate the method by considering the quadratic Li\'{e}nard type nonlinear system  described classically  by the equation of motion, 
\begin{eqnarray}
\ddot{x} + f(x) \dot{x}^2 + h(x) = 0,
\label{lie}
\end{eqnarray}
where $f(x)$ and $h(x)$ are arbitrary functions. The corresponding Hamiltonian reads \cite{pmm, tiwari} as
\begin{eqnarray} 
H = \frac{p^2}{2 m(x)} + V(x), \label{H} 
\end{eqnarray}
where the mass, $m(x) =  e^{2\int^x  f(x') dx'}$ and the momentum, $p = m(x) \dot{x}$. The potential, $V(x)$ is of the form 
\begin{eqnarray}
V = \int^{x} m(x') h(x') dx'.  \label{pote}
\end{eqnarray}

Recently, the above nonlinear ordinary differential equation (\ref{lie}) has been investigated for its Lie point symmetry properties \cite{tiwari}. The general form of 	Eq. (\ref{lie}) 
has been classified based on the fact whether the equation admits one, two, three or eight parameter Lie point symmetry groups. The general form of (\ref{lie}) which admits maximal eight parameter symmetry group, and is also linearizable through local transformations, is of the form 
\begin{eqnarray}
\ddot{x} + f(x) \dot{x}^2 + 2 \lambda_1 e^{-\int^x f(x') dx'} \int^x  e^{-\int^{x'} f(x'') dx''} dx'+ 2 \lambda_2  e^{-\int^x f(x') dx'} = 0, \label{g1} 
\end{eqnarray}
where $\lambda_1$ and $\lambda_2$ are constants. One can show that the corresponding Hamiltonian \cite{ml_ham} is 
\begin{eqnarray}
\hspace{-1.2cm}\qquad H_1 = \frac{p^2}{2}  e^{-2\int^x f(x') dx'} + \lambda_1 \left(\int^{x} e^{\int^{x'} f(x'') dx''} dx'\right)^{2} 
+ 2 \lambda_2 \int^{x} e^{\int^{x'} f(x'') dx''} dx'. \label{h1} 
\end{eqnarray}
On the other hand the particular  form of  (\ref{lie}), namely  
\begin{eqnarray}
\ddot{x} + f(x) \dot{x}^2 - \frac{\lambda_3}{2}  e^{-\int^x f(x') dx'} \left(\int^x  e^{\int^{x'} f(x'') dx''} dx'\right)^{-3} +2\lambda_1 e^{-\int^x f(x') dx'} \int^x  e^{\int^{x'} f(x'') dx''} dx' = 0, \nonumber \\ 
\label{g2}
\end{eqnarray}
is integrable but linearizable through nonlocal transformations, corresponding to 
three parameter symmetry groups \cite{tiwari}.  The corresponding Hamiltonian \cite{ml_ham} is 
\begin{eqnarray}
H_2 = \frac{p^2}{2}e^{-2\int^x f(x') dx'} + \frac{\lambda_3}{4} \left(\int^{x} e^{\int^{x'} f(x'') dx''} dx'\right)^{-2} + \lambda_1 \left(\int^{x} e^{\int^{x'} f(x'') dx''} dx'\right)^2. 
\label{h2}
\end{eqnarray}
It has been also shown that the systems (\ref{g1}) and (\ref{g2}) represent the isochoronous oscillations if the relevant functions of (\ref{lie}) satisfy the conditions \cite{tiwari}, 
\begin{eqnarray}
\hspace{-0.3cm}\mbox{(i)}\; h' + f h &=& 2 \lambda_1, \quad \left(' = \frac{d}{d x}\right),\label{iso1}\\
\hspace{-0.3cm}\mbox{(ii)}\; h' + f h &=& 2\lambda_1  + \frac{3}{2} \lambda_3 \left(\int^x e^{\int^{x'} f(x'') dx''} dt\right)^{-4},  \label{iso2}
\end{eqnarray}
respectively. 

The above relations on using (\ref{pote}) yield two different potentials, namely 
\begin{eqnarray}
\mbox{(i)}\;V_1(x) \equiv U_1(g) &=&  \lambda_1 g^2(x) + 2 \lambda_2 g(x),\label{pot1}\\
\mbox{(ii)}\;V_2(x) \equiv U_2(g) &=& \frac{\lambda_3}{4\;g^2(x)} + \lambda_1 g^2(x),   \label{pot2}
\end{eqnarray}
where $g(x) = \int^{x} \sqrt{m(x')} d x'$ and $m(x) =  e^{2\int^x  f(x') dx'}$, corresponding to the above Hamiltonians $H_1$ and $H_2$ respectively (see Eqs. (\ref{h1}) and (\ref{h2})). 

We solve the corresponding position dependent mass Hamiltonians for the above 
two potentials based on the method discussed above. 

\subsection{\bf Case 1 Potential $V_1(x)$}
The potential
\begin{eqnarray}
V_1(x) \equiv U_1(g) = \lambda_1 g^2(x) + 2 \lambda_2 g(x),\nonumber \hspace{1cm}(\ref{pot1})
\end{eqnarray}
is in the form of a linear harmonic oscillator potential. For this potential, we consider both 
Hermitian and non-Hermitian ordered forms of quantum Hamiltonian corresponding to its 
classical counterpart (\ref{h1}) and implement the above procedure for these two cases. 

\subsubsection{\bf (a)\;Hermitian ordering}

The corresponding one-dimensional Hermitian Hamiltonian (\ref{geham_heo}) for the potential $V_1(x)$ is   
\begin{eqnarray}
\hat{H}_{her} = \frac{1}{2}\hat{p}\frac{1}{m}\hat{p} + \frac{\hbar^2}{2}\left[\left(\frac{\bar{\alpha}+\bar{\gamma}}{2}\right) \frac{d^2}{d x^2}\left(\frac{1}{m}\right)+ \left(\overline{\alpha\gamma}   \frac{1}{4} (\bar{\gamma} - \bar{\alpha})^2\right)\left(\frac{d}{dx}\left(\frac{1}{m}\right)\right)^2 m\right]+ V_1 (x),  \nonumber \\
\label{geham1}
\end{eqnarray}
where $V_1(x) \equiv  U_1(g(x))$ is as given in Eq. (\ref{pot1}). 
The associated time-independent Schr\"{o}dinger equation is 
\begin{eqnarray}
\frac{-\hbar^2}{2 m}\left[\psi{''} - \frac{m{'}}{m} \psi{'} +  
\left(\left(\frac{\bar{\alpha}+\bar{\gamma}}{2}\right)  \frac{m{''}}{m}-\left(\overline{\alpha\gamma}+\bar{\gamma}+\bar{\alpha} +\frac{1}{4}(\bar{\gamma} - \bar{\alpha})^2\right) \frac{m{'^2}}{m^2}\right)\right]\psi\nonumber \\+ (\lambda_1 g^2(x) + 2 \lambda_2 g(x))\psi 
= E \psi. 
\label{seg1aa}
\end{eqnarray}
Here the prime stands for differentiation with respect to $g(x)$. 
By implementing the procedure discussed earlier, we can transform Eq. (\ref{seg1aa}) for the potential $V_1(x)$ to the constant mass Schr\"{o}dinger equation (\ref{seg_phi3_he}). To prove this, we first introduce the PCT (\ref{pct}) along with (\ref{gm}) for $d = \frac{1}{4}$ which reduces Eq. (\ref{seg1aa}) to an equation for $\phi(g)$ as   
\begin{eqnarray}
\hspace{-0.3cm}\;\frac{d^2 \phi}{d g^2} + \left[\left(\bar{\alpha} + \bar{\gamma} + \frac{1}{2}\right) \frac{g'''}{g'^3} 
-\left(\bar{\alpha}^2 + \bar{\gamma}^2 -  2 \bar{\alpha} \bar{\gamma} + 3 \bar{\alpha} +3 \bar{\gamma}  + 4 \overline{\alpha\gamma} +\frac{5}{4}\right) \frac{g''^2}{g'^4} \right. \nonumber \\
\hspace{-0.3cm}\; \left.+ \frac{2}{\hbar^2} (E - \lambda_1 g^2(x) - 2 \lambda_2 g(x)) \right] \phi = 0. \label{seg_phi11}
\end{eqnarray}
We then remove the terms involving the derivatives of $g$ through two ways by treating either mass or ordering parameters as arbitrary as discussed in the previous sections and obtain 
\begin{equation}
\frac{d^2 \phi}{d g^2} + \frac{2}{\hbar^2} (E - \lambda_1 g^2(x) - 2 \lambda_2 g(x)) \phi = 0. 
\label{seg_phi11}
\end{equation}
An asymptotic analysis of (\ref{seg_phi11}) as $g \rightarrow \infty$ suggests the transformation 
\begin{equation}
\phi(g) = \exp{\left[-\frac{1}{\hbar \sqrt{2\;\lambda_1}} \left(\lambda_1\;g^2 + 2 \lambda_2\;g\right)\right]}\;\chi(g),
\end{equation}
which leads to the equation
\begin{equation}
 \frac{d^2 \chi}{d g^2} -\frac{2 \sqrt{2}}{\hbar \sqrt{\lambda_1}} \left(\lambda_1 g +\lambda_2 \right)  \frac{d \chi}{d g} + \left[ \frac{2 E}{\hbar^2} + \frac{2 \lambda_2^2}{\hbar^2 \lambda_1} - \frac{\sqrt{2 \lambda_1}}{\hbar} \right] \chi = 0.
\label{gpeq11a}
\end{equation}

Now introducing the  transformation,
\begin{equation}
\tau = \left(\frac{\sqrt{2 \lambda_1}}{\hbar}\right)^{1/2} \left(g + \frac{\lambda_2}{\lambda_1}\right), \label{ta}
\end{equation}
which reduces  (\ref{gpeq11a}) to the form of the Hermite differential equation,
\begin{equation}
\frac{d^2 \chi}{d \tau^2} -2 \tau \frac{d \chi}{d \tau}
+  \frac{\hbar}{\sqrt{2 \lambda_1}}\left( \frac{2 E}{\hbar^2} + \frac{2  \lambda_2^2}{\hbar^2 \lambda_1}-  \frac{\sqrt{2 \lambda_1}}{\hbar} \right) \chi = 0, 
\label{chih}
\end{equation}
provided ${\displaystyle \frac{\hbar}{\sqrt{2 \lambda_1}}\left( \frac{2 E_n}{\hbar^2} + \frac{2  \lambda_2^2}{\hbar^2 \lambda_1}-  \frac{\sqrt{2 \lambda_1}}{\hbar} \right)= 2 n}, \; n = 0, 1, 2, 3, ...$. Hence, we can  obtain the  energy eigenvalues and eigenfunctions of (\ref{geham1}) as
\begin{eqnarray}
E_n &=& (2 n + 1) \hbar \sqrt{\frac{\lambda_1}{2}} - \frac{\lambda^2_2}{\lambda_1}, \label{en1} \\
\psi_n(x) &=& N_n \exp{\left[-\frac{1}{\hbar \sqrt{2 \lambda_1}} \left(\lambda_1\;g^2 + 2 \lambda_2\;g\right)\right]} m^{\frac{1}{4}}(x)  \times H_n\left[\left(\frac{\sqrt{2 \lambda_1}}{\hbar}\right)^{1/2} \left(g + \frac{\lambda_2}{\lambda_1} \right)\right],  \nonumber \\
& & \hspace{8cm}g(x) = \int^{x} \sqrt{m(x')} d x', \label{eig1}
\end{eqnarray}
where $H_n(y), \; n =0, 1, 2, 3, ...,$ are Hermite polynomials \cite{book}. 

The normalization constant $N_n$ can be obtained using the normalization condition as follows:
\begin{eqnarray}
1 &=& \langle \psi_n | \psi_n\rangle \nonumber \\
 &=& N^2_n \int^{\infty}_{-\infty} \exp{\left[-\frac{\sqrt{2}}{\hbar \sqrt{\lambda_1}} \left(\lambda_1\;g^2 + 2 \lambda_2\;g\right)\right]} m^{\frac{1}{2}}(x) \left(H_n\left[\left(\frac{\sqrt{2 \lambda_1}}{\hbar}\right)^{1/2} \left(g + \frac{\lambda_2}{\lambda_1} \right)\right]\right)^2 dx. \nonumber \\
 \label{norm1}
\end{eqnarray}
Since $g' = \sqrt{ m(x)}$, from Eq. (\ref{norm1}), we obtain 
\small
\begin{eqnarray}
1 &=& N^2_n \int^{\infty}_{-\infty} \exp{\left[-\frac{\sqrt{2}}{\hbar \sqrt{\lambda_1}} \left(\lambda_1\;g^2 + 2 \lambda_2\;g\right)\right]}  \left(H_n\left[\left(\frac{\sqrt{2 \lambda_1}}{\hbar}\right)^{1/2} \left(g + \frac{\lambda_2}{\lambda_1} \right)\right]\right)^2 d g, 
\end{eqnarray}
 \normalsize
which reduces to 
\begin{eqnarray}
1 = N^2_n \frac{\sqrt{\hbar}}{(2 \lambda_1)^{1/4}} \exp{\left(\sqrt{\frac{2}{\lambda_1}}\frac{\lambda_2^2}{\hbar \lambda_1}\right)}\int^{\infty}_{-\infty} e^{-\tau^2} H_n(\tau) H_n(\tau) d \tau,   \label{norm2}
\end{eqnarray}
by applying the transformation (\ref{ta}). Since ${\displaystyle \int^{\infty}_{-\infty} e^{-\tau^2} H_n(\tau) H_n(\tau) d \tau = 2^n n! \sqrt{\pi}}$ \cite{book}, from Eq. (\ref{norm2}) we can obtain the normalization constant 
as  
\begin{equation}
N_n = \left(\frac{\exp{\left(-\frac{\lambda^2_2 }{\hbar\;\lambda_1}\sqrt{\frac{2 }{\lambda_1}}\right)} (2 \lambda_1)^{\frac{1}{4}}}{\sqrt{\hbar \pi} 2^n n!}\right)^{\frac{1}{2}}. \label{norm_he_pot1}
\end{equation}

Now we analyse the results for the two cases, that is (i) mass $(m(x))$ is arbitrary and (ii) the ordering parameters $\alpha_i, \beta_i, \gamma_i,\;i = 1, 2, 3,...N$ are arbitrary, separately. 

{\bf Sub-case (i) $\bf m(x)$ as arbitrary}

As discussed in the section \ref{case1_he}, when mass function is considered to be arbitrary, only one Hermitian ordering given by (\ref{her_he}) or (\ref{geham1}) is possible. 

Following the analysis given above,  we can conclude that the position dependent mass counterparts of $U_1(g)$ corresponding to arbitrary  mass functions are exactly solvable if the associated Hamiltonians are ordered as  in (\ref{geham1}).  The solutions (\ref{eig1}) can be explicitly expressed as   
 \begin{eqnarray}
\hspace{-1cm}E_n &=& (2 n + 1) \hbar \sqrt{\frac{\lambda_1}{2}} - \frac{\lambda^2_2}{\lambda_1}, \nonumber \\
\hspace{-1cm}\psi_n(x) &=& N_n m^{\frac{1}{4}}(x) \exp{\left[-\frac{1}{\hbar}\sqrt{\frac{\lambda_1}{2}}\;\left(\int^x \sqrt{m(x')} dx'\right)^2  - \frac{\lambda_2}{\hbar}\sqrt{\frac{2}{\lambda_1}}\;\int^x \sqrt{m(x')} dx'\right]} \nonumber \\ 
\hspace{-0.2cm} & & \times H_n\left[\left(\frac{\sqrt{2 \lambda_1}}{\hbar}\right)^{1/2} 
\left(\int^x \sqrt{m(x')} dx' + \frac{\lambda_2}{\lambda_1} \right)\right],
\end{eqnarray}
where $N_n$ are the normalization constants obtained in (\ref{norm_he_pot1}).  

{\bf Sub-case (ii) ordering parameters as arbitrary}

As discussed in \ref{case2_he}, when the ordering parameters are considered to be arbitrary, 
the functions $g(x)$ and $m(x)$ are explicitly fixed to be (\ref{g}) and (\ref{massess}), respectively. 
Correspondingly the potential $V_1(x)$ (\ref{pot1}) takes particular forms 
\small
\begin{subnumcases}{\label{eq:pot1_he2} \hspace{-0.6cm}V_1(x) = }
  \!\!\lambda_1 \mu^2_1 e^{2 \mu_2 x} + 2 (\lambda_2 - \lambda_1 \mu_3) \mu_1 e^{\mu_2 x}
            +\lambda_1 \mu^2_3 - 2 \lambda_2 \mu_3,\label{eq:a}\\
   \!\! \lambda_1  \left(\nu_1 x + \nu_2\right)^{2 c} + 2 (\lambda_2 - \lambda_1 \mu_3) 
         (\nu_1 x + \nu_2)^c   + \lambda_1 \mu^2_3 - 2 \lambda_2 \mu_3, \;-\infty < c < \infty, \label{eq:b}
  \end{subnumcases}
\normalsize
where $\mu_1, \; \mu_2, \; \mu_3, \nu_1, \nu_2$ and $c$ are arbitrary parameters. 
The above class of position dependent mass potentials which are also counterparts to $U_1(g)$, are explicitly solvable for arbitrary  choices of ordering parameters, $\alpha_i, \beta_i, \gamma_i$, $i = 1, 2, 3, ...N$.  The associated Hermitian ordered form of Hamiltonian (\ref{geham_heo}) of the potentials (\ref{eq:pot1_he2}) admit eigenfunctions (vide (\ref{eig1})), respectively,  as 
\begin{subnumcases}{\label{eig1_pot1}\psi_n(x)= }
N_n \exp{\left[-\frac{1}{\hbar \sqrt{2 \lambda_1}} \left(\lambda_1 \mu^2_1\; e^{2 \mu_2 x} + 2 (\lambda_2 - \lambda_1 \mu_3)\;\mu_1\; e^{\mu_2 x}  + \lambda_1 \mu^2_3 - 2 \lambda_2 \mu_3\right)\right]} a^{1/4}_1\;e^{a_2 x/2}\nonumber \\
\hspace{2cm}\times    H_n\left[\left(\frac{\sqrt{2 \lambda_1}}{\hbar}\right)^{1/2} \left(\mu_1 e^{\mu_2 x} - \mu_3 + \frac{\lambda_2}{\lambda_1} \right)\right], \\
  N_n \exp{\left[\lambda_1 \left(\nu_1 x + \nu_2 \right)^{2 c} + 2 (\lambda_2 - \lambda_1 \mu_3)  \left(\nu_1 x + \nu_2\right)^{c} + \lambda_1 \mu^2_3 - 2 \lambda_2 \mu_3\right]} \left(b_1 x + b_2\right)^{\frac{C}{4}}\nonumber\\
\hspace{2cm} \times  H_n\left[\left(\frac{\sqrt{2 \lambda_1}}{\hbar}\right)^{1/2} \left(\left(\nu_1 x + \nu_2\right)^{c} - \mu_3+ \frac{\lambda_2}{\lambda_1} \right)\right],  -\infty < c < \infty, 
 \end{subnumcases}
\normalsize  
where $N_n$, $n = 0, 1, 2, 3, ...$ are the normalization constants obtained in (\ref{norm_he_pot1}), for the energy eigenvalues $E_n$ (\ref{en1}). The parameters $\mu_1, \; \mu_2, \; \mu_3$, $\nu_1, \nu_2$, $c$ and $C$ are defined in Eqs. (\ref{gess}) and 
(\ref{massess}). 

\subsubsection{\bf (b) Non-Hermitian ordering} 
Similarly while considering the non-Hermitian Hamiltonian (\ref{geham_non}) for the one-dimensional potential $V_1(x)$, 
\begin{eqnarray}
\hat{H}_{non} &=& \frac{1}{2}\hat{p}\frac{1}{m}\hat{p}+(\bar{\gamma} - \bar{\alpha}) \frac{i \hbar}{2} \frac{d}{dx}\left(\frac{1}{m}\right) \hat{p} + \frac{\hbar^2}{2}\left[\bar{\gamma} \frac{d^2}{dx^2}\left(\frac{1}{m}\right)  + \overline{\alpha\gamma} \left(\frac{d}{dx}\left(\frac{1}{m}\right)\right)^2 m \right] + V_1(x),     \nonumber \\
\label{geham_non_pot2}
\end{eqnarray}
it can be related with the Hermitian Hamiltonian (\ref{geham1}) through the transformation (\ref{heta}). Hence, from the solutions (\ref{eig1}) of the Hermitian Hamiltonian, we can obtain the eigenfunctions of the non-Hermitian Hamiltonian (\ref{geham_non_pot2}) along with the energy eigenvalues, ${\displaystyle E_n = (2 n + 1) \hbar \sqrt{\frac{\lambda_1}{2}} - \frac{\lambda^2_2}{\lambda_1}}$, as 
\begin{eqnarray}
\hspace{-2cm} \qquad \quad \;\tilde{\psi}_n(x) &=& N_n \exp{\left[-\frac{1}{\hbar \sqrt{2 \lambda_1}} \left(\lambda_1\;g^2 + 2 \lambda_2\;g\right)\right]} m^{\frac{1}{2}\left(\bar{\alpha} - \bar{\gamma} + \frac{1}{2}\right)} H_n\left[\left(\frac{\sqrt{2 \lambda_1}}{\hbar}\right)^{1/2} \left(g + \frac{\lambda_2}{\lambda_1} \right)\right]. \label{eig2}
\end{eqnarray} 

Now we analyze the results for the two cases, that is (i) mass $(m(x))$ is arbitrary and (ii) the ordering parameters $\alpha_i, \beta_i, \gamma_i,\;i = 1, 2, 3,...N$ are arbitrary.

{\bf Sub-case (i) $\bf m(x)$ as arbitrary}

In this case the $2 N$ ordering parameters of the non-Hermitian Hamiltonian (\ref{geham_non}) get  reduced to one parameter ($\gamma$) corresponding to the  Hamiltonian for the potential $V_1(x)$, that is 
\begin{eqnarray}
\hat{H}_{non} = \frac{1}{2} m^{-\frac{1}{2}-\gamma} \hat{p} m^{-\frac{1}{2}} \hat{p} m^{\gamma}+ V_1(x),   
\label{non_pot1}
\end{eqnarray}
which can be re-expressed as 
\begin{eqnarray}
\hat{H}_{non} &=& \frac{1}{2}\hat{p}\frac{1}{m}\hat{p}+  {i \hbar} \left({\gamma}+ \frac{1}{4}\right) \frac{d}{dx}\left(\frac{1}{m}\right) \hat{p} +\frac{\hbar^2}{2} \left[\gamma \frac{d^2}{dx^2}\left(\frac{1}{m}\right) -\gamma\left(\gamma + \frac{1}{2}\right) \left(\frac{d}{dx}\left(\frac{1}{m}\right)\right)^2 m \right]+ V_1(x). \nonumber \\
\label{geham1_non_pot1}
\end{eqnarray}
From (\ref{eig2}), the solutions for the system (\ref{geham1_non_pot1}) can be obtained as   
 \begin{eqnarray}
E_n &=& (2 n + 1) \hbar \sqrt{\frac{\lambda_1}{2}} - \frac{\lambda^2_2}{\lambda_1}, \nonumber \\
\tilde{\psi}_n(x) &=& N_n \exp{\left[-\frac{1}{\hbar}\sqrt{\frac{\lambda_1}{2}}\;\left(\int^x \sqrt{m(x')} dx'\right)^2 - \frac{\lambda_2}{\hbar}\sqrt{\frac{2}{\lambda_1}}\;\int^x \sqrt{m(x')} dx'\right]} 
m^{-\gamma}(x) \nonumber \\
 & & \times H_n\left[\left(\frac{\sqrt{2 \lambda_1}}{\hbar}\right)^{1/2} 
\left(\int^x \sqrt{m(x')} dx' + \frac{\lambda_2}{\lambda_1} \right)\right], \nonumber
\end{eqnarray}
where $N_n$ are the normalization constants obtained in (\ref{norm_he_pot1}).  

{\bf Sub-case (ii) ordering parameters as arbitrary}

As discussed earlier in the Hermitian ordering, the functions $g(x)$ and $m(x)$ take specific forms (\ref{gess}) and (\ref{massess}) which fix the potential $V_1(x)$ to be particular forms (\ref{eq:pot1_he2}). The non-Hermitian Hamiltonian for the two classes of potentials (\ref{eq:pot1_he2}) are   
\small
\begin{eqnarray}
\hat{H}_{non} &=& \frac{1}{2}\hat{p}\frac{1}{m}\hat{p}+(\bar{\gamma} - \bar{\alpha}) \frac{i \hbar}{2} \frac{d}{dx}\left(\frac{1}{m}\right) \hat{p} + \frac{\hbar^2}{2}\left[\bar{\gamma} \frac{d^2}{dx^2}\left(\frac{1}{m}\right)  + \overline{\alpha\gamma} \left(\frac{d}{dx}\left(\frac{1}{m}\right)\right)^2 m \right]  + \lambda_1 \mu^2_3 - 2 \lambda_2 \mu_3 \nonumber \\ 
& &+ 2 (\lambda_2 - \lambda_1 \mu_3) \mu_1\; e^{\mu_2 x} +  \lambda_1 \mu^2_1\; e^{\left(2 \mu_2 x\right)}, \\
\mbox{and} & & \nonumber \\
\hat{H}_{non} &=& \frac{1}{2}\hat{p}\frac{1}{m}\hat{p}+(\bar{\gamma} - \bar{\alpha}) \frac{i \hbar}{2} \frac{d}{dx}\left(\frac{1}{m}\right) \hat{p} + \frac{\hbar^2}{2}\left[\bar{\gamma} \frac{d^2}{dx^2}\left(\frac{1}{m}\right)  + \overline{\alpha\gamma} \left(\frac{d}{dx}\left(\frac{1}{m}\right)\right)^2 m \right]+  \lambda_1  \left(\nu_1 x + \nu_2\right)^{2 c} \nonumber \\ 
& &+ 2 (\lambda_2 - \lambda_1 \mu_3) (\nu_1 x + \nu_2)^c + \lambda_1 \mu^2_3 - 2 \lambda_2 \mu_3, \;\;\;-\infty < c < \infty.   
\label{geham_pot1_or}
\end{eqnarray}
\normalsize 
The corresponding time independent Schr\"{o}dinger equation can be exactly solved  and the solutions can be written using (\ref{eig2}) as
\begin{subnumcases}{\label{eig1_pot1} \tilde{\psi}_n(x) =}
 N_n\;\exp{\left[-\frac{1}{\hbar \sqrt{2 \lambda_1}} \left(\lambda_1 \mu^2_1\; e^{2 \mu_2 x} + 2 (\lambda_2 - \lambda_1 \mu_3)\;\mu_1 e^{\mu_2 x} + \lambda_1 \mu^2_3 - 2 \lambda_2 \mu_3\right)\right]} a^{-\gamma}_1\;e^{-\gamma\;a_2 x} \nonumber \\
\hspace{2cm}\times  H_n\left[\left(\frac{\sqrt{2 \lambda_1}}{\hbar}\right)^{1/2} \left(\mu_1 e^{\mu_2 x} - \mu_3 + \frac{\lambda_2}{\lambda_1} \right)\right], 
\\
N_n \exp{\left[\lambda_1 \left(\nu_1 x + \nu_2 \right)^{2 c} + 2 (\lambda_2 - \lambda_1 \mu_3)  \left(\nu_1 x + \nu_2\right)^{c}  + \lambda_1 \mu^2_3 - 2 \lambda_2 \mu_3\right]}\left(b_1 x + b_2\right)^{-\gamma\;C} \nonumber\\
\hspace{2cm}\times   H_n\left[\left(\frac{\sqrt{2 \lambda_1}}{\hbar}\right)^{1/2} \left(\left(\nu_1 x + \nu_2\right)^{c} - \mu_3+ \frac{\lambda_2}{\lambda_1} \right)\right], 
\;-\infty < c < \infty, 
\end{subnumcases}
where $N_n$, $n = 0, 1, 2, 3, ...$ are the normalization constants obtained in (\ref{norm_he_pot1}), for the energy eigenvalues $E_n$ (\ref{en1}). 

\subsection{\bf Case 2 Potential $V_2(x)$}
Now we consider the inverse square form of the potential,
\begin{eqnarray}
\hspace{1cm}V_2(x) \equiv V_2(g) = \frac{\lambda_3}{4\;g^2(x)} +\lambda_1 g^2(x). \nonumber \hspace{1cm}(\ref{pot2})
\end{eqnarray}

{\bf (i) Hermitian ordering}

Using the above form of $V_2(x)$, the Schr\"{o}dinger equation corresponding to Hermitian ordered form of Hamiltonian can be written as 
\begin{eqnarray}
\frac{-\hbar^2}{2 m}\left[\psi{''} - \frac{m{'}}{m} \psi{'} +  
\left(\left(\frac{\bar{\alpha}+\bar{\gamma}}{2}\right)  \frac{m{''}}{m}-\left(\overline{\alpha\gamma}+\bar{\gamma}+\bar{\alpha} +\frac{1}{4}(\bar{\gamma} - \bar{\alpha})^2\right) \frac{m{'^2}}{m^2}\right)\right]\psi\nonumber \\ 
+ \left(\frac{\lambda_3}{4\;g^2(x)} +\lambda_1 g^2(x)\right)\psi= E \psi. 
 \label{seg1a}
\end{eqnarray}
Following the same procedure discussed for the potential $V_1$, Eq. (\ref{seg1a}) can be transformed to the constant mass equation 
\begin{eqnarray}
\frac{d^2 \phi}{d g^2} + \left[\frac{2}{\hbar^2} \left(E -\frac{\lambda_3}{4\;g^2(x)} -\lambda_1 g^2(x)\right) \right] \phi &=& 0. 
\label{seg_phi1b}
\end{eqnarray}
This is of the form of an isotonic oscillator \cite{calogero}. Hence, 
the energy eigenvalues and eigenfunctions of the system (\ref{pot2}) from (\ref{sol}) read as 
\begin{eqnarray}
\hspace{-0.3cm}\qquad E_n &=&  2 \hbar \sqrt{2 \lambda_1} \left( n + \frac{l}{2} + \frac{3}{4}\right) , \\
\hspace{-0.3cm}\qquad \psi_n(x) &=& \psi_{n,+} = N_n \; \exp{\left[-\sqrt{\frac{\lambda_1}{2\;\hbar^2}}\;g^2(x)\right]} m^{\frac{1}{4}} g^{l + 1}(x)L^{l + \frac{1}{2}}_n\left[ \frac{\sqrt{2\;\lambda_1}}{\hbar} g^2(x) \right], \; 0 < g < \infty, \nonumber \\
\hspace{-05.cm}\qquad \psi_{n,-}(x) &=& \pm \psi_{n,+}(-g), \; \; -\infty < g < 0,  \label{iso_g_s}
\end{eqnarray}
where the normalization constant $N_n$ reads as 
\begin{equation}
N_n = \left[2 \left(\frac{\sqrt{2 \lambda_1}}{\hbar}\right)^{l + \frac{3}{2}}\frac{n!}{
\Gamma{(n + l + \frac{3}{2})}}\right]^{\frac{1}{2}}. \label{norn}  
\end{equation}
Here $L^{l + \frac{1}{2}}_n(y),\;n=0,1,2,3,...$ are the associated Laguerre polynomials of degree $n$ and order $l+\frac{1}{2}$ \cite{book} and ${\displaystyle \frac{1}{2}\sqrt{1+ \frac{2\sqrt{2 \lambda_3}}{\hbar^2}}-\frac{1}{2}= l}$, where $l+\frac{1}{2}$ is an integer. 

{\bf (ii) Non-Hermitian ordering}

Similarly while considering the non-Hermitian Hamiltonian (\ref{geham_non}) for a one-dimensional potential $V_2(x)$, 
\begin{eqnarray}
\hat{H}_{non} &=& \frac{1}{2}\hat{p}\frac{1}{m}\hat{p}+(\bar{\gamma} - \bar{\alpha}) \frac{i \hbar}{2} \frac{d}{dx}\left(\frac{1}{m}\right) \hat{p} + \frac{\hbar^2}{2}\left[\bar{\gamma} \frac{d^2}{dx^2}\left(\frac{1}{m}\right)  + \overline{\alpha\gamma} \left(\frac{d}{dx}\left(\frac{1}{m}\right)\right)^2 m \right] + V_2(x), \nonumber \\
\label{geham_non_pot2_2}
\end{eqnarray}
the energy eigenvalues and eigenfunctions can be obtained  from (\ref{sol}) using (\ref{heta})  
\begin{eqnarray}
\hspace{-1cm}\qquad E_n &=&  2 \hbar \sqrt{2 \lambda_1} \left( n + \frac{l}{2} + \frac{3}{4}\right) , \\
\hspace{-1cm}\qquad \tilde{\psi}_n(x) &=& \psi_{n,+} = N_n \; \exp{\left[-\sqrt{\frac{\lambda_1}{2\;\hbar^2}}\;g^2(x)\right]} m^{\frac{1}{2}\left(\bar{\alpha} - \bar{\gamma} + \frac{1}{2}\right)} 
g^{l + 1}(x) L^{l + \frac{1}{2}}_n\left[ \frac{\sqrt{2\;\lambda_1}}{\hbar} g^2(x) \right], \; 0 < g < \infty, \nonumber \\
\hspace{-1cm}\qquad \tilde{\psi}_{n,-}(x) &=& \pm \tilde{\psi}_{n,+}(-g), \; \; -\infty < g < 0,  \label{iso_g_s}
\end{eqnarray}
where the normalization constant, $N_n$, obtained in (\ref{norn}).

Then as in the case of the potential $V_1(x)$, we can consider the two subcases, namely (i) $m(x)$ arbitrary and (ii) ordering parameters arbitrary, using the appropriate forms of $g(x)$ and $m(x)$. We do not present them explicitly here as their forms are obvious from the previous discussion. 

\section{Conclusion}
We have considered a general Hermitian ordered form of  kinetic energy operator corresponding to a quantum particle with PDM. Using point canonical transformation method the associated generalized Schr\"{o}dinger equation is transformed to a constant mass Schr\"{o}dinger equation endowed with a generalized potential considered to be exactly solvable and so the solutions of the PDM systems can be obtained. Hence, we have obtained a 
class of exactly solvable position-dependent mass counterparts to each constant mass potential.  The solutions of PDM systems preserving Hermiticity do not depend on a particular ordering for a certain class of mass functions. It is also remarkable to state that for certain specific choices of the mass function, the proposed method of solving removes the ambiguity in the problem of ordering. We have pointed out the methods of relating the Hermitian counterpart with its non-Hermitian Hamiltonian and obtained eigenfunctions of the non-Hermitian Hamiltonian. They include the ordering parameters and so the ordering ambiguity 
continues to be present in the case of non-Hermitian ordering. We illustrated the above results with  two generalized nonlinear oscillators of  quadratic Li\'{e}nard type nonlinear differential equations obtained using Lie point symmetry technique.

\section{Acknowledgments}
VC wishes to thank the Council of Scientific and Industrial Research,
Government of India, for providing a Senior Research Fellowship. The work forms
a part of a research project of MS, and an IRHPA project and a Ramanna Fellowship project
of ML, sponsored by the Department of Science and Technology (DST), Government of India. ML
also acknowledges the financial support under a DAE Raja Ramanna Fellowship.

\end{document}